%% file: article_3.tex
\documentclass[fleqn,10pt]{SelfArx} 

\usepackage{longtable}
\usepackage{xurl}
\usepackage{hyperref}
\usepackage{tocloft}
\usepackage[page]{appendix}
\usepackage{pdfpages}
\usepackage{multirow}

\cftsetindents{section}{0em}{2em}
\cftsetindents{subsection}{0em}{2em}

\definecolor{color1}{RGB}{0,0,90} 
\definecolor{color2}{RGB}{0,20,20} 

\hypersetup{hidelinks,colorlinks,breaklinks=true,urlcolor=color2,citecolor=color1,linkcolor=color1,bookmarksopen=false,pdftitle={Title},pdfauthor={Author}}

\JournalInfo{.} 
\Archive{} 

\PaperTitle{Malicious Source Code Detection Using Transformer}

\Authors{Chen Tsfaty\textsuperscript{1}* and Michael Fire\textsuperscript{2}*}
\affiliation{\textsuperscript{1}\textit{chents@post.bgu.ac.il}} 
\affiliation{\textsuperscript{2}\textit{mickyfi@bgu.ac.il}} 
\affiliation{*Department of Software and Information Systems Engineering, Ben-Gurion University} 

\newcommand{\keywordname}{Keywords} 

\begin{document}
\Abstract{\input{abstract}}

\maketitle 
\thispagestyle{empty}

\section{Introduction} 

\input{introduction}


\section{Related Work}
\input{related_work}


\section{Methods}
\input{methods}


\section{Results}
\input{results}


\section{Discussions}
\input{discussions}


\section{Conclusions and Future Works}
\input{conclusion_and_future_work}

\section{Data and Code Availability}
\input{data_and_code_availability}

\phantomsection

\phantomsection
\bibliographystyle{unsrt}
\bibliography{mybib}


\input{appendix}

\end{document}

%% file: abstract.tex
Open source code is considered a common practice in modern software development. However, reusing other code allows bad actors to access a wide developers' community, hence the products that rely on it. Those attacks are categorized as supply chain attacks. Recent years saw a growing number of supply chain attacks that leverage open source during software development, relaying the download and installation procedures, whether automatic or manual. Over the years, many approaches have been invented for detecting vulnerable packages. However, it is uncommon to detect malicious code within packages. Those detection approaches can be broadly categorized as analyzes that use (dynamic) and do not use (static) code execution. Here, we introduce \textit{M}alicious \textit{S}ource code \textit{D}etection using \textit{T}ransformers (MSDT) algorithm. MSDT is a novel static analysis based on a deep learning method that detects real-world code injection cases to source code packages. In this study, we used MSDT and a dataset with over 600,000 different functions to embed various functions and applied a clustering algorithm to the resulting vectors, detecting the malicious functions by detecting the outliers. We evaluated MSDT's performance by conducting extensive experiments and demonstrated that our algorithm is capable of detecting functions that were injected with malicious code with \textit{precision@k} values of up to 0.909.

%% file: introduction.tex
\label{chap:intro}

Software \textit{supply chain attacks} aim to access source codes, build processes, or update mechanisms by infecting legitimate apps to distribute malware.\footnote{\url{https://docs.microsoft.com/en-us/windows/security/threat-protection/intelligence/supply-chain-malware/}} Hence the end-users will refer to that malware as trusted software, e.g., download or update sites. An illustrative example of such attacks is the Codecov attack \cite{jackson_2021}, a backdoor concealed within a Codecov uploader script that is downloaded vastly. In April 2021, attackers compromised a Codecov server to inject their malicious code into a bash uploader script. Codecov customers then downloaded this script for two months. When executed, the script exfiltrated sensitive information, including keys, tokens, and credentials from those customers' Continuous Integration/ Continuous Delivery (CI/CD) environments. By utilizing this data, Codecov attackers reportedly breached hundreds of customer networks, including HashiCorp, Twilio, Rapid7, Monday.com, and e-commerce giant Mercari~\cite{jackson_2021}. 

Those types of attacks are becoming more popular and harmful \cite{sonatype2021state} due to modern development procedures. Those procedures use open-source packages and public repositories for many reasons: efficiency, accelerating development, cost-effectiveness, etc. For that reason, open-source demand is becoming widespread among many developers. With a 73\% growth of components downloaded in 2021 compared to 2020 \cite{sonatype2021state}. The development procedures that involve those packages and repositories are mostly automatic, such as build procedures or semi-automatic, the same as developers installing an open-source package~\cite{chris2021beware}. As a result of the mentioned growth, popular packages, development communities, lead contributors, and many more can be considered attractive targets for \textit{software supply chain attacks} \cite{NIST2021defending,sawers2021next,sharma2021newly,peterson2021software,gregory2021supply}. That kind of attack may pass their vulnerability to dependent software projects. By 2021, OWASP considers \textit{software supply chain threat} one of the Top-10 security issues worldwide.\footnote{\url{https://owasp.org/www-project-top-ten/}} A lead example of such attacks is \textit{ua-parser-js} attack \cite{sharma_2021}, which occurred in October 2021. The attacker was granted ownership of the package by account takeover and published three malicious versions. At that time, \textit{ua-parser-js} was a highly popular package with more than seven million weekly downloads. 

In recent years, a vast research field has emerged to issue with this threat \cite{NIST2021defending,ohm2020backstabber}. This field is researched by academia and is part of the application security market, which was valued at 6.42 billion USD \cite{marketsandmarkets2020application}. This field includes many aspects that depend on various parameters, such as (1) programming language (PL). For example, different PLs have different security issues \cite{georgian2020common, kelly2021cpp}; and (2) the scope of examining functionalities (function, class, scripts, etc.). For example, there are attacks targeting a centric function \cite{bertus2019discord} or modules \cite{constantin2018npm}.

\begin{figure*}[ht]\centering
\includegraphics[width=\linewidth]{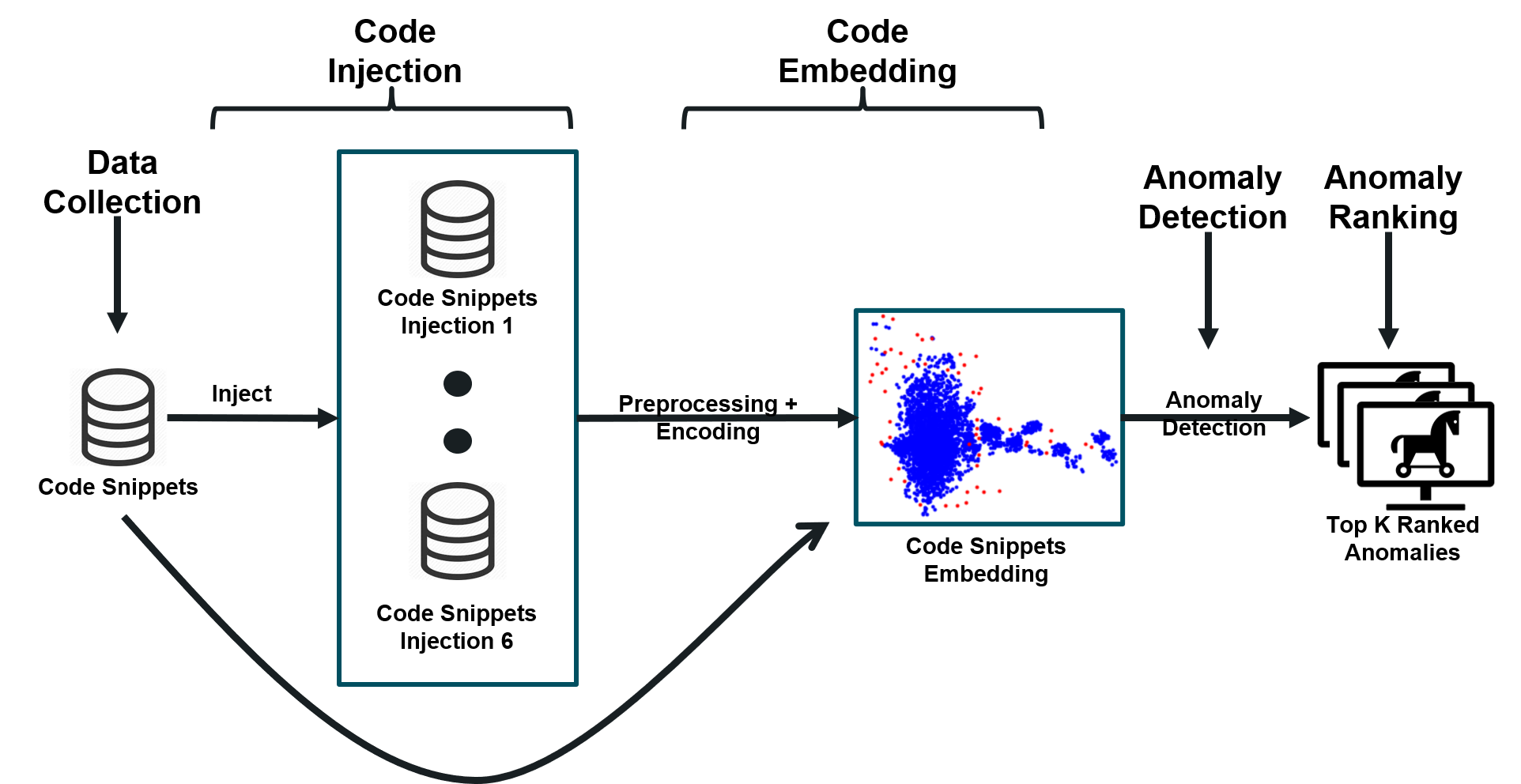}
\caption{\label{fig:method} Overview of our data embedding and anomaly detection model process.}
\end{figure*}

In this study, we developed the MSDT algorithm, a novel method for detecting malicious code injection within functions' source code, by static analysis that consists of the following four key steps (see Figure \ref{fig:method} and Section \ref{sec:proposed_method}): First, we used the \textit{PY150} dataset \cite{raychev2016probabilistic} to train a transformer architecture model. Second, by utilizing the transformer, we were able to embed every function in the \textit{CodeSearchNet (CSN) Python} dataset, which is used for experiments evaluation, \cite{husain2019codesearchnet} into the representation space of the transformers' encoding part. Third, we applied a clustering algorithm over every function type implementation to detect anomalies by outlier research. Lastly, we ranked the anomalies by their distance from the nearest clusters' border points - the farther the point is, the higher the score.

We conducted extensive experiments to evaluate MSDT's performance. The experiments concluded, randomly injecting to the top 100 common functions five different real-world malicious codes, Code2Seq \cite{alon2018code2seq} as the transformer, and DBSCAN for  the clustering algorithm \cite{prado2017dbscan}.
Eventually, we evaluate the results by \textit{precision at $k$ (precision@k)} (for various $k$ values) of matching functions classified as malicious with their true tagging (see Section \ref{sec:conducted_experiments}). The \textit{precision@k} test result values measured by applying MSDT reached up to 0.909. For example, MSDT achieved this result when $k=20$ for the different implementations of the \textit{get} function. Those implementations were randomly injected with a real-world attack presented by Bertus et al. \cite{bertus2019discord}. Additionally, we empirically evaluated MSDT on a real-world attack and succeeded in detecting it. Lastly, we empirically compared MSDT to widely used static analysis tools, which are only able to work on files, while MSDT works on functions. MSDT's capability to work on functions gives a more precise ability to detect an injection in a given function.

The key contributions of our study are threefold:
\begin{enumerate}
\item We have developed MSDT, a novel algorithm to automatically detect code injection via anomaly detection in functions' source code.
\item We have created MSDT to support any textual PL. We can ensure it by using the proper grammar and a transformer architecture (Code2Seq \cite{alon2018code2seq}) to embed functions' source code.
\item We have curated an open dataset of 607,461 functions that were injected with several real-world malicious codes. This dataset can be used in future works in the field of detection code injections.
\end{enumerate}

The remainder of the paper is structured as follows: Section \ref{chap:related_work} summarizes the related work. Section \ref{chap:methods_and_experiments} describes the proposed methodology and the conducted experiments in the study. Section \ref{chap:results} presents the results of this study. That is followed by Section \ref{chap:discussions}, in which we discuss the study results. Lastly, Section \ref{chap:conclusion} summarizes and concludes the study and offers future work.

%% file: related_work.tex
\label{chap:related_work}
Malformed open-source packages constitute several threats to every component in some development procedures and have become a vast research field with three main branches \cite{ogasawara1998experiences}. In the following subsection, we provide an overview of these branches: Section \ref{sec:security_issues} introduces an overview of the security issues that commonly appear in public repositories or occur due to the PL features weaknesses exploitation. Next, Section \ref{sec:detection_methods} provides an overview of the widely used methods to detect those attacks or weaknesses. Lastly, Section \ref{sec:deep_learning_methods} gives an overview of the different Deep Learning (DL) methods in the field of code representation, which are used to apply advanced static analysis to the targeted code.

\subsection{Security issues within open source packages}
\label{sec:security_issues}
In recent years, the awareness of the threats regarding public repositories and open-source packages has increased. As a result, many studies \cite{ohm2020backstabber, Harush2021, birsan2021dependency} point out two main security issues with the usage of those packages: (1) \textit{vulnerable packages} \cite{snyk2021opencv}- which contain a flaw in their design \cite{snyk2018double}, unhandled code error \cite{snyk2020unchecked} or other bad practices that could be a future security risk \cite{ruohonen2021large, ruohonen2018empirical}. This threat is widespread and has been vastly researched by communities or commercial companies (e.g., Snyk\footnote{\url{https://snyk.io/}} and WhiteSource\footnote{\url{https://www.whitesourcesoftware.com/}}). Usually, this threat is based on Common Vulnerabilities and Exposures (CVEs).\footnote{\url{https://cve.mitre.org/}}$^{,}$\footnote{\url{https://snyk.io/vuln}} Those vulnerabilities allow the malicious actor, with prior knowledge of the package usage location, to achieve its goal with a few actions \cite{Tal2020, Sharma2021}; and (2) \textit{malicious intent in packages} \cite{tschacher2016typosquatting}- which contain bad design, unhandled code error, piece of code that is not serving the main functionality of the program, etc. Those examples are created to be exploited or triggered at some phases of the package (installation, test, runtime, etc.).

Studies have shown a rise in malicious functionalities appearing in public repositories and highly used packages \cite{ruohonen2021large, zimmermann2019small, polkovnichenko_2022}. With this rise, it becomes clear that there are common injection methods for malicious actors to infect packages. As demonstrated by Ohm et al. \cite{ohm2020backstabber}, to inject malicious code into a package, an attacker may either infect an existing package or create a new package that will be similar to the original one (often called \textit{dependency confusion} \cite{birsan2021dependency}). A new malicious package developed and published by a malicious actor has to follow several principles: (1) To make a proper replacement to the targeted package, it has to contain a semi-ident functionality; and (2) It has to be attractive, ending up in the targeted users’ dependency tree. To grant the use of those new packages types, one of the following methods can suit: Naming the malicious package similar to the original one (\textit{typosquatting}) \cite{bertus2019discord,birsan2021dependency,tschacher2016typosquatting,cimpanu2018twelve}, creating a \textit{trojan} in the package \cite{constantin2018npm, cimpanu2019malicious}, using an unmaintained package, or user account (\textit{use after free}) \cite{claburn2018resurrect}. As mentioned, the second injection strategy is to infect existing packages in one of the following methods: (1) Inject to the source of the original package, by a Pull request / social engineering \cite{chris2021beware, della2021anatomy, thomas2018compromised, us2021malware}; (2) The open source project owner added malicious functionality out of ideology, such as political \cite{paganini2022nodeipc}; (3) Inject during the build process \cite{kisielius2021breaking}; and (4) Inject through the repositories system \cite{cappos2008look}.

Ohm et al. \cite{ohm2020backstabber} demonstrated that the malicious intent in packages could be categorized by several parameters: targeted OS (Operating System), PL, the actual malicious activity, the location of the malicious functionality within the package (where it is injected), and more. Additionally, they showed the majority of the maliciousness is associated with persistence purposes, which can be categorized into several major groups: Backdoors, Droppers, and Data Exfiltration \cite{ohm2020backstabber}.

In this study, we focus on the second security issue with a specification in a dynamic PL (\textit{Python} as a test case) for the reasons of usage popularity and the popularity of injection-oriented attacks within those PLs repositories (\textit{Node.js}, \textit{Python}, etc.) \cite{ohm2020backstabber}. Those injections are often related to the PLs dynamicity features \cite{georgian2020common}, such as exposing the running functionalities only at runtime (e.g., exec(“print (Hello world!)”)), configurable dependencies and imports of packages (e.g., import from a local package instead of a global one).

The described use of the PLS dynamicity features is the most common among the known attacks \cite{ohm2020backstabber,sonatype2020state}. A leading example of this kind of attack was presented by Bertus~\cite{bertus2019discord}. Bertus reviewed a malicious package named ``pytz3-dev,'' which was seen in PyPI\footnote{\textit{Python package index} - the main repository of Python packages} and downloaded by many. This package contains malicious code in the initialization module and searches for a Discord authentication token stored in an SQLite database. Eventually, the code exfiltrated the token if found. This attack was carried out unnoticed for seven months and downloaded by 3000 users in 3 months \cite{bertus2019discord, sonatype2020state}. Those features, and many more, are used by attackers, making this threat one of the most common attack techniques associated with a \textit{supply chain attack}, as covered by NIST \cite{NIST2021defending}.  

\subsection{Detection methods of malicious intent in source code}
\label{sec:detection_methods}
As a result of the increase in the mentioned above security issues, two major detection methods were developed:

\subsubsection{Static Analysis}
\label{sec:static_analysis}
A type of analysis that finds irregularities in a program without executing it. The irregularities can broadly be categorized into three main branches: coding style enforcement, reliability, and maintainability \cite{ruohonen2021large, lizdenis2020configure}. The security issues are mainly associated with the reliability domain, which primarily covers bug detection~\cite{wang2010detect}, vulnerability detection~\cite{russell2018automated}, and malware detection challenges~\cite{idika2007survey, patil2017detection}. To deal with those challenges, the following are common techniques in static analysis that gather information regarding the detection mission:
\begin{itemize}
\item \textit{Syntax properties}. This technique uses the PL syntax to find irregularities. For example, using AST to search obfuscated strings that are most likely to be executed \cite{bertus2018detecting} or a linter operation to check the program's correctness \cite{lizdenis2020configure}. 
\item \textit{Feature-based technique}. This technique uses the occurrences count of known problematic functionalities \cite{ruohonen2021large, garrett2019detecting}. For example, Patil et al. \cite{patil2017detection} have constructed a classifier with a given labeled dataset and several features extracted (function appearances, length of the script, etc.) that can predict the maliciousness of a script. The main drawback of this technique is that it strongly binds with reversing research that points to features related to the attack, which may lead to detection overfitting the attacks that have been revealed and learned. Secondly, potential attackers could evade detection by several methods, such as not using or properly using the searched features in the code \cite{flesman2019evading}.

An example of such a static analysis tool is Bandit \cite{bandit_2022}. Bandit is a widespread tool \cite{ruohonen2021large} designed to find common security issues in Python files, using hard-coded rules. This tool uses AST (see Section \ref{sec:deep_learning_methods}) form of the source code to better examine the rule set. In addition, Bandit detection method includes the following metrics: \textit{severity} of the issues detected and the \textit{confidence} of detection for a given issue. Those metrics are divided into three values: \textit{low}, \textit{medium} and \textit{high}. Each rule gets its \textit{severity} and \textit{confidence} values manually by Bandits' community.

\item \textit{Data preprocess}. Construct a workable data structure that grasps the syntax and semantic information of the code to represent the code better (see Section \ref{sec:deep_learning_methods}). It will be convenient to apply anomaly detection or classification research with a proper code representation. For example, Alomari et al. \cite{alomari2019scalable} construct a control flow graph, and by resemblance subgraphs, they manage to identify similar code segments between programs.
\item \textit{Signature-based detection} (in the case of malware detection) is a process where a set of rules (based on reversing procedure) define the maliciousness level of the program \cite{sentinelone2021what}. Those rules that are generated for static analysis purposes are often a set of functionalities or opcodes in a specific order to match the researched code behavior. For example, YARA\footnote{\url{https://github.com/Yara-Rules/rules}} is a commonly used static signature tool; and the rules that are generated for dynamic analysis purposes are often a set of executed operations, memory states, registers' values, etc. \cite{idika2007survey}. The main drawback of this technique is that it applies to known maliciousness.
\item \textit{Comparing packages} to known CVEs (see Section \ref{sec:security_issues}).
\end{itemize}
On the one hand, static analysis tends to scale well over many PL classes (with a given grammar), efficiently operates on large corpora, often will identify well-known security issues, and in many cases, is explainable \cite{pvs2015static}. On the other hand, this kind of analysis suffers from a high number of false positives and poor configuration issues detection \cite{wichers2020source}.

\subsubsection{Dynamic Analysis}
\label{sec:dynamic_analysis}
Those type of analysis is a group that finds irregularities in a program after its execution and determines its maliciousness. In this type of analysis, the gathered data (system calls, variable values, IO access, etc.) are often used as part of anomaly detection or classification problem \cite{idika2007survey}. There are several drawbacks for this type of analysis on a source code \cite{pvs2013dynamic}: (a) \textit{Data gathering difficulties}- there is a need to activate the package and execute its functionality, hence making the procedure of extracting data hard to automate; and (b) \textit{Scalability} - there is a need to activate all the learned and tested program, and for each to extract the wanted data. In this study, we will focus on advanced static analysis.

\subsection{Deep learning methods for analyzing source code}
\label{sec:deep_learning_methods}
In recent years, there has been an increasing need to use machine learning (ML) methods in code intelligence for productivity and security improvement \cite{lu2021codexglue}. As a result, many studies construct statistical models to code intelligence tasks. Recently, pre-trained models were constructed by learning from big PL corpora, such as CodeBERT \cite{feng2020codebert} and CodeX \cite{chen2021evaluating}. These pre-trained models are commonly based on models from the natural language process (NLP) field, such as BERT \cite{devlin2018bert} and GPT \cite{brown2020language}. This development led not only to improvement in code understanding \cite{lu2021codexglue} and generation problems \cite{alon2020structural} but also to enlarging the number of tasks and their necessity \cite{lu2021codexglue}, such as Clone detection \cite{ain2019systematic} and Code completion \cite{raychev2014code}.
Those tasks include several challenges, such as capturing semantic essence \cite{nagar2021code}, syntax resemblance \cite{alomari2019scalable}, and figure execution flow \cite{yu2019empirical}. For every challenge, it occurred that there is a model that will fit better than others \cite{lu2021codexglue}. For example, for code translating between PLs, algorithms that include a “Cross-lingual Language Model'' with masked tokens preprocessing are superior for capturing the semantic essence well \cite{feng2020codebert,lachaux2020unsupervised}.

Over the years, several ML methods have been researched in the context of code analysis tasks. In 2012, Hovsepyan et al. \cite{hovsepyan2012software} showed the use of techniques from the classic text analysis field, for example, using SVM on a bag-of-words (BOW) representation of simple tokenization (lexing by the PL grammar) of Java source. In 2016, Dam et al. \cite{dam2016deep} and Liang et al. \cite{liang2018automatic} presented techniques to get context for the extracted tokens, for example, using the output of recurrent neural network (RNN) trained over tokenized (lexing representations) code \cite{dam2016deep}. However, according to Ahmad et al. \cite{ahmad2020transformer}, RNN-based sequence models lack several source code concepts regarding source code representations. First, inaccurate representation of the non-sequential structure of source code. Second, RNN-based models may be inefficient for very long sequences. Third, those models lack to grasp of the syntactic and semantic information of the source code. Therefore, starting in 2018, studies include two significant changes in learning source code representation. First is the use of \textit{Transformers}, which have proven to be efficient in capturing long-range dependencies~\cite{alon2020structural}. Second are the different data preprocessing procedures, which yields more informatically data structures to learn on: Alon et al. \cite{alon2018code2seq} used AST Paths for a transformer architecture named Code2Seq \cite{alon2018code2seq}, Mou et al. \cite{mou2014tbcnn} utilized \textit{abstract syntax tree}\footnote{Abstract Syntax Tree (AST) is a well-known data structure for representing a program with a given PL grammar (see \url{https://www.twilio.com/blog/abstract-syntax-trees} for further explanation).} nodes to train tree-based convolutional neural networks for supervised classification problems. Lately, researchers have tried to include semantic data of the PLs. For example, Feng et al. \cite{feng2020codebert} presented the CodeBERT model, which uses a bimodal pre-trained model to learn the semantic relationship between natural language and PLs such as Java, PHP, Python, etc.

\begin{figure}
\centering
\includegraphics[width=0.3\textwidth]{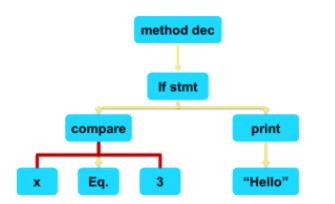}
\caption{\label{fig:AST} Example AST transformation of the code snippet if x == 3: print(“Hello”). Example of AST path painted in red.}
\end{figure}

In this study, we used the Code2Seq model, a transformer architecture developed by Alon et al. \cite{alon2018code2seq}. Additionally, similarly to Ramakrishnan et al. \cite{ramakrishnan2020semantic}, we trained the model using the PY150 dataset \cite{mou2014tbcnn} - a dataset that contains Python functions in the form of AST (see Section \ref{sec:datasets}). In this architecture, a function is referred to as an AST. The output trees' internal nodes represent the construction of the program with known rules, as described in the given grammar. The tree’s leaves represent information regarding the program variables, such as names, types, values, etc. Figure \ref{fig:AST} outlines the notion of AST on code snippets.

Eventually, the Code2Seq model gets as an input a set of AST paths\footnote{Every pairwise path between two leaf tokens is represented as a sequence containing the AST nodes. Those nodes are connected by up and down arrows. These arrows exemplify the up or downlink between the nodes in the tree. Example for an AST path that is shown in Figure \ref{fig:AST} : (x, ↑if stmt, ↑method dec ↓print: “Hello”).} that were extracted from code snippets. A bi-directional LSTM encodes those paths to create a vector representation for each path and its AST values separately.
Then the decoder attends over those encoded paths while generating the target sequence. The final output of the Code2Seq model is generated sequence of words that explain the functionality of the given code snippet \cite{alon2018code2seq}.

Code2seq can be integrated into many applications \cite{alon2018code2seq, nagar2021code, ramakrishnan2020semantic}, such as code search - with a given sentence describing a code, and the output will be the wanted code. For example, Nagar et al. \cite{nagar2021code} used the Code2seq model to generate comments for collected code snippets. Then, the candidate code snippets and corresponding machine-generated comments are stored in a database. Eventually, the code snippets whose comments are semantically similar to natural language queries are retrieved.

Recent studies have presented more advanced code embedding methods that try to include the program's semantic, syntactic, and execution flow as part of the representation \cite{alomari2019scalable, yu2019empirical}.

%% file: methods.tex
\label{chap:methods_and_experiments}
The primary goal of this study is to detect code injection by applying static analysis to the source code. This section describes the static analysis algorithm we developed (see Section \ref{sec:proposed_method}) and our experiments to test and evaluate our proposed method, MSDT (see Section \ref{sec:conducted_experiments}).

\subsection{The proposed method}
\label{sec:proposed_method}
As presented in Section \ref{sec:security_issues}, in supply chain attacks, the injected functionality will often be added to the source of the targeted program. Therefore, the code will be changed. This study presents MSDT, an algorithm to detect the mentioned difference in the program’s functionality for a chosen PL, by the four following steps (see Figure \ref{fig:method}):
\begin{enumerate}
\item \textit{Data collection}. In this step, we collect a sufficient amount of function implementations of the chosen PL, for each function type. For example, to detect code injection in the "encode" function, we collect a sufficient amount of "encode" implementations to better estimate the distribution of the implementations. In addition, the collected data can be different versions of the same function. The collection of data can be manually collected from any code-base warehouse (such as GitHub) or extracted from an existing code dataset. For example, an existing dataset of functions with their names and implementations  (see Section \ref{sec:datasets}).

\item \textit{Code embedding}. In this step, we create an embedding layer to the given source code snippets by using an algorithm that gets sequence data and represents it as a vector. An example of such algorithms is transformers that vectorize the input sequence and transform it to another sequence, such as Seq2seq \cite{ramakrishnan2020semantic}, Code2seq \cite{alon2018code2seq}, CodeBERT \cite{feng2020codebert}, and TransCoder \cite{lachaux2020unsupervised}. The resulting embedding layer has to be reasonable so that similarity in the source code snippets (similar functions) translates to a similarity in the embedding space. For example, the vectors of the square-root and cube-root functions will be relatively close to each other and farther than the parse timezone function's vector.

\item \textit{Anomaly detection}. In this step, we apply an anomaly detection technique by applying cluster algorithms and detecting the outliers. For example, we can utilize DBSCAN and K-means to cluster the input and detect outliers \cite{badr_2019}. We use this technique on every function type embedding layer and manage to differentiate code snippets that were injected from benign code snippets.

\item \textit{Anomaly ranking}. Lastly, we rank the outliers by their distance from the nearest clusters’ border points in this step \cite{huang2013rank}. The farther the point is, the higher the score.
\end{enumerate}

\subsection{Experiments}
\label{sec:conducted_experiments}
There are several datasets including labeled function implementations for several purposes \cite{lu2021codexglue}. In this study, we used 607,461 public \textit{Python} function implementations, with simulated test cases and real-world observed attacks. Additionally, this study combines an embedding layer based on a transformer, Code2Seq \cite{alon2018code2seq}. Lastly, this study showcases traditional anomaly detection techniques over the Code2Seq representation based on DBSCAN \cite{prado2017dbscan} compared to another anomaly detection technique based on Ecod \cite{li2022ecod}.

\subsubsection{Datasets}
\label{sec:datasets}
In this study, we utilized three datasets: (1) The PY150 dataset \cite{raychev2016probabilistic} is used for training Code2Seq. The PY150 is a Python corpus with 150,000 files. Each file contains up to 30,000 AST nodes from open-source projects with non-viral licenses such as MIT. For the training procedure, we randomly sampled the PY150 dataset to validation/test/train sets of 10K/20K/120K files; (2) The CodeSearchNet (CSN) \textit{Python} dataset \cite{husain2019codesearchnet} is used for evaluating the different experiments. CSN is a \textit{Python} corpus, containing 457,461 $<docstring, code>$ pairs from open source libraries, which we refer only to as the code; and (3) The Backstabber’s Knife Collection \cite{ohm2020backstabber} is used for the malicious functionalities injected during the simulations. The Backstabber’s Knife Collection is a dataset of manual analysis of malicious code from 174 packages that were used by real-world attackers. 
Namely, we use five different malicious code injections from this collection, to inject in the 100 most common functions within the CSN corpus. We chose those specific malicious codes for their straightforward integration within the injected function, and their download popularity \cite{ohm2020backstabber}.

As mentioned above, the input to the Code2seq model is an AST representation of a function. To get this representation for each function, we extracted tokens using \textit{fissix}\footnote{\url{https://github.com/jreese/fissix}} and \textit{tree\_sitter},\footnote{\url{https://github.com/tree-sitter/tree-sitter}} which allows us to normalize the code to get consistent encoding. With the normalized output code, we then generate an AST using \textit{fissix}.

\subsubsection{Injection simulation}
\label{sec:injection_simulation}
To simulate the real-world number of code injections, we randomly selected up to 10\% \cite{sonatype2021state} implementations from each of the top 100 common functions to be code injected,\footnote{To find the 100 most common functions we count the number of implementations for each function in the CSN dataset, and refer to the 100 most frequent functions.} with a total of 48627 implementations. The injected functionalities were five malicious samples collected from \textit{Backstabber’s Knife Collection} \cite{ohm2020backstabber}. Those injections illustrate several attacks types:
\begin{enumerate}
\item \textit{A one-liner execution of obfuscated string}, encoded by base64 \cite{bertus2019discord}. This string is a script that finds the Discord chat application’s data folder on Windows machines and then attempts to extract the Discord token from an SQLite database file. Once the Discord token is found, it is sent to a web server.\footnote{We use two different execution functions (in different types of injections), \textit{exec} and \textit{os.system} functions. These functions allow the user to execute a string.}
\item \textit{A one-liner execution of non-obfuscated script} - the deobfuscation of the described above attack.
\item \textit{Loading a file from the root directory of the program}. The loaded file is a keylogger that eventually sends the collected data to a remote server via emails. To mask the keylogger loading, we are using the Popen function to execute the malicious functionality in other subprocesses \cite{meyers_tozer_2020}.
\item \textit{Attacker payload construction as an obfuscation use case}.\footnote{\url{https://securityboulevard.com/2020/08/string-concatenation-obfuscation-techniques/}} We splitted the obfuscated string (the first attack mentioned in this section) into several substrings. Then we concatenate those strings in several parts of the program to construct the original attacker string.\footnote{\label{refnote} Executing the concatenate string using \textit{os.system} function.}
\end{enumerate}
The injected functionalities were injected at the beginning of the randomly selected implementations for those popular function types, similar to the mentioned attacks above~\cite{bertus2019discord, meyers_tozer_2020} and as viewed by Ohm et al.~\cite{ohm2020backstabber}.

\subsubsection{Code2seq representation}
\label{sec:code2seq_representation}
In this study, we use the result vectors of the attention procedure (see Section \ref{sec:deep_learning_methods}), named \textit{context vectors} with 320 dimensions - it is the representation space of the model for code snippets. At each decoding step, the probability of the next target token depends on the previous tokens \cite{alon2018code2seq}.

We used Alon et al. \cite{alon2018code2seq} implementation for Code2Seq\footnote{\url{https://github.com/tech-srl/code2seq }} model and set it with the same parameters. We trained the Code2Seq model on a server with a high RAM setting.\footnote{The server specifications are: 256G RAM and 48 CPU cores. The training process continued for 24 hours on 130K functions.} We construct the encoder to be two bi-directional LSTMs that encode the AST paths consisting of 128 units each, and we set a dropout of 0.5 on each LSTM. Then, we construct the decoder to be an LSTM consisting of one layer with size 320, and we set a dropout of 0.75 to support the generation of longer target sequences. At last, we trained the model for 20 epochs or until there was no improvement after 10 iterations. Eventually, we test our Code2seq model on the PY150 test set (as mentioned in Section~\ref{sec:datasets}) and achieved the following metrics on the mentioned randomly sampled test set: recall of 47\%, precision of 64\%, and F1 of 54\%.

\subsubsection{Anomaly detection on representation}
\label{sec:anomaly_detection_on_representation}
In this step, we use our Code2Seq representation (see Section \ref{sec:code2seq_representation}) for the given injected functions and non-injected from the same type. Then, we test several clustering algorithms, such as DBSCAN, K-means, Ecod, and Hierarchical clustering. Eventually, we chose the DBSCAN method (referred to as $MSDT_{DBSCAN}$) to find outliers because it works well on multi-dimensional data, as presented by Oskolkov et al.~\cite{Nikolay_Oskolkov_2019}.
We achieved it by using tuning the following parameters for the DBSCAN method \cite{prado2017dbscan}:
\begin{enumerate}
\item\textit{eps} which specifies the distance between two points, and is testing with the following values: 0.2 - 1.0.
\item\textit{min\_samples} which specifies the minimum number of neighbors to consider a point in a cluster, and is testing with the following values: 2 - 10.
\end{enumerate}
For each iteration, we apply 10-fold cross-validation and measure the following metrics by the mean of the different folds: TPR, AP (Average Precision), and detecting outlier precision.

\subsubsection{Evaluation Process}
\label{sec:evaluation_metrices}
The performance of the anomalies detected by MSDT was measured by precision at $k$ (\textit{precision@k}) study, which stands for the true positive rate (TPR) of the results that occurs within the top $k$ of the ranking~\cite{ruohonen2021large}. We rank the anomalies by their Euclidean distance from the nearest clusters’ border points. Eventually, we measured the \textit{precision@k} metric for each function type with the mentioned code injection attacks and compared it to a $Random Classifier$, to show the performance of MSDT relatively to a random decision. Additionally, to understand better the way MSDT detects attacks, we examine the correlation between the detection rate and the number of implementations among the various function types. Therefore we measured the average \textit{precision@k} for every attack, and for every function type, we calculated the average of the average detection rate of the various attacks. We used Spearman's rank correlation ($\rho$) to measure the correlation between the mentioned average of the function types and their number of implementations. 

We compared $MSDT_{DBSCAN}$’s performance to another widely use outlier detection baseline method name \textit{Ecod} (referred to as $MSDT_{Ecod}$) \cite{li2022ecod} over the mentioned representation (see Section \ref{sec:anomaly_detection_on_representation}). We use Ecod to detect outliers as follows: First, we apply Ecod on every function type for every attack type (accordingly to $MSDT_{DBSCAN}$). Second, we measure the anomaly score of each implementation.\footnote{The Ecod algorithm calculates this score. The more the vector is distant, the higher its score.} Third, we extract the \textit{precision@k} where $k$ indicates the anomalies in descending order, i.e, \textit{precision@2} is the precision of the two most highly ranked anomalies, as simulated by Amidon et al.~\cite{amidon2022ecod}.

To evaluate our method on real-world injections, we applied $MSDT_{DBSCAN}$ on a real-world case taken from the Backstabber’s Knife Collection \cite{ohm2020backstabber}. The case is a sample of malicious functionality injected in \textit{multiply} calculation functionality that loads a file by Popen, as mentioned above in Section ~\ref{sec:injection_simulation}. We collected 48 implementations of \textit{multiply} relate functions from the mentioned datasets (see Section \ref{sec:datasets}). We did so to gain reference of the injected \textit{multiply} function to the benign implementations, and thus we were able to apply $MSDT_{DBSCAN}$ on this \textit{multiply} case.

Additionally, we compared MSDT with the mentioned $MSDT_{Ecod}$ method and two of the well-known static analysis tools named \textit{Bandit} and \textit{Snyk} (see Section \ref{sec:static_analysis}). Namely, we evaluate those static analysis tools on the origin file where the malicious implementation of \textit{multiply} appeared.

Lastly, to emphasize the relations between the malicious and the benign implementations, we visualized the achieved embedding of the \textit{get} and the \textit{log} functions with the injected code. We manage this visualization by applying PCA (2 components) \cite{li2019pca} on the Code2Seq context vectors (see Section \ref{sec:code2seq_representation}).

%% file: results.tex
\label{chap:results}

In this section, we present the experimental results, which were obtained by the MSDT algorithm (see Section \ref{sec:proposed_method}) when applied to the constructed function types dataset that contains both injected and benign implementations (see Section \ref{sec:injection_simulation}).\footnote{We utilize 8G RAM with 8 CPU cores server to evaluate the algorithm. The runtime of the process took 10 minutes for 48627 different implementations.} 

\begin{figure*}[!h]
\includegraphics[width=\linewidth]{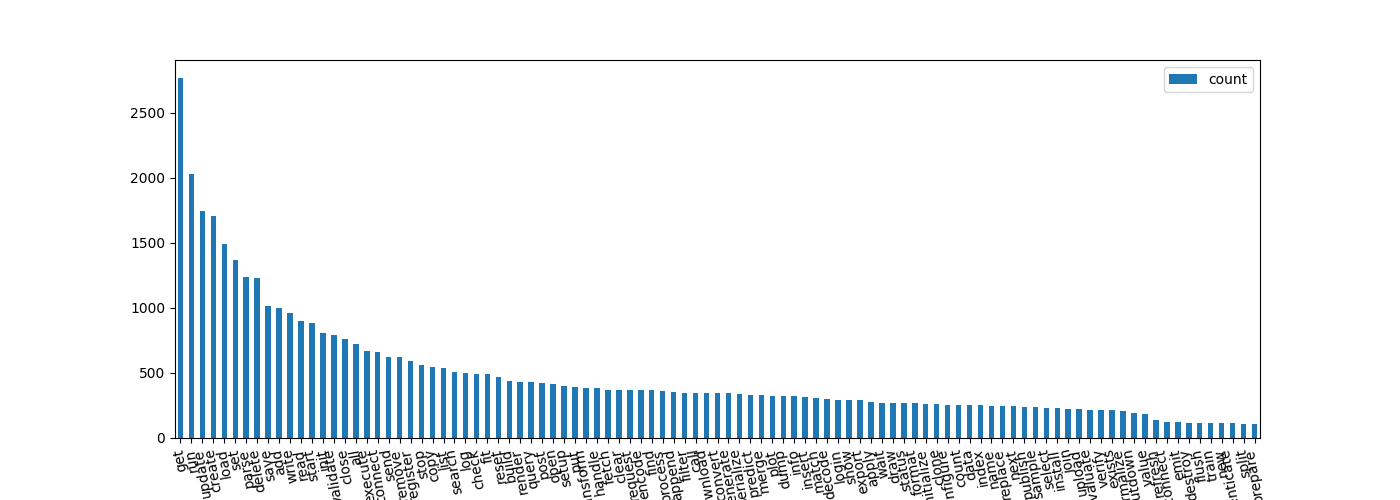}
\caption{\label{fig:imp_per_func} Number of different implementations per functions' types.}
\end{figure*}

The constructed dataset includes the 100 most common function types from the CSN dataset (see Section \ref{sec:datasets}). From the function types implementations distribution (see Figure \ref{fig:imp_per_func}), the most common function type is the \textit{get} function with over of 3,000 unique implementations; and the least common from those function types is the \textit{prepare} with 102 unique implementations.

\begin{figure*}[!h]
\includegraphics[width=\linewidth]{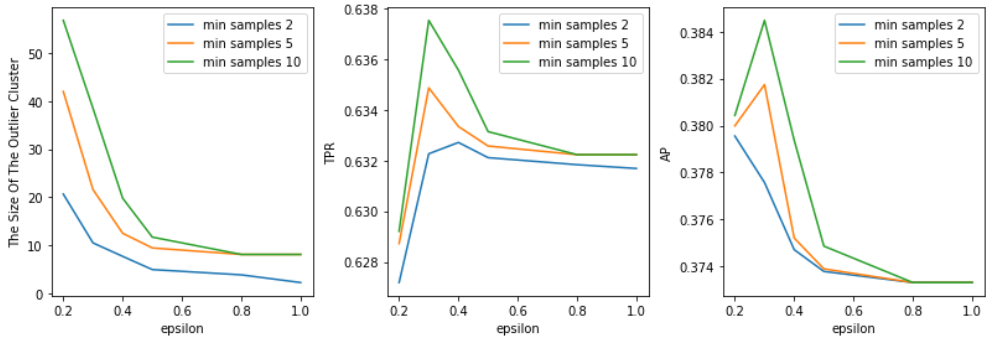}
\caption{\label{fig:dbscan_parameter} The following graphs show the DBSCAN parameter tuning process: (1) The size of the outlier cluster, that indicate whether the methods overfit or underfit; (2) The measured \textit{precision@k} for a range of $k$; and (3) The measured AP (average precision) for a range of $k$.}
\end{figure*}

The first experiment included parameter tuning of the DBSCAN method mentioned in Section~\ref{sec:anomaly_detection_on_representation}. We received the following best results (see Figure~\ref{fig:dbscan_parameter}) for \textit{eps}=0.3 and \textit{min\_samples}=10: \textit{TPR}=0.637, \textit{AP}=0.384, detecting outlier precision=0.953. These results indicate that it is possible to detect anomalies by finding outliers with probable rates. 
In addition, when the default values of the DBSCAN method is set \cite{schubert2017dbscan}, we got \textit{TPR}=0.632, \textit{AP}=0.373, detecting outlier precision=0.738. Therefore, the DBSCAN with the tuned parameters exceeded the one with the default parameters.

\begin{figure*}
\includegraphics[width=1.1\textwidth,height=12cm]{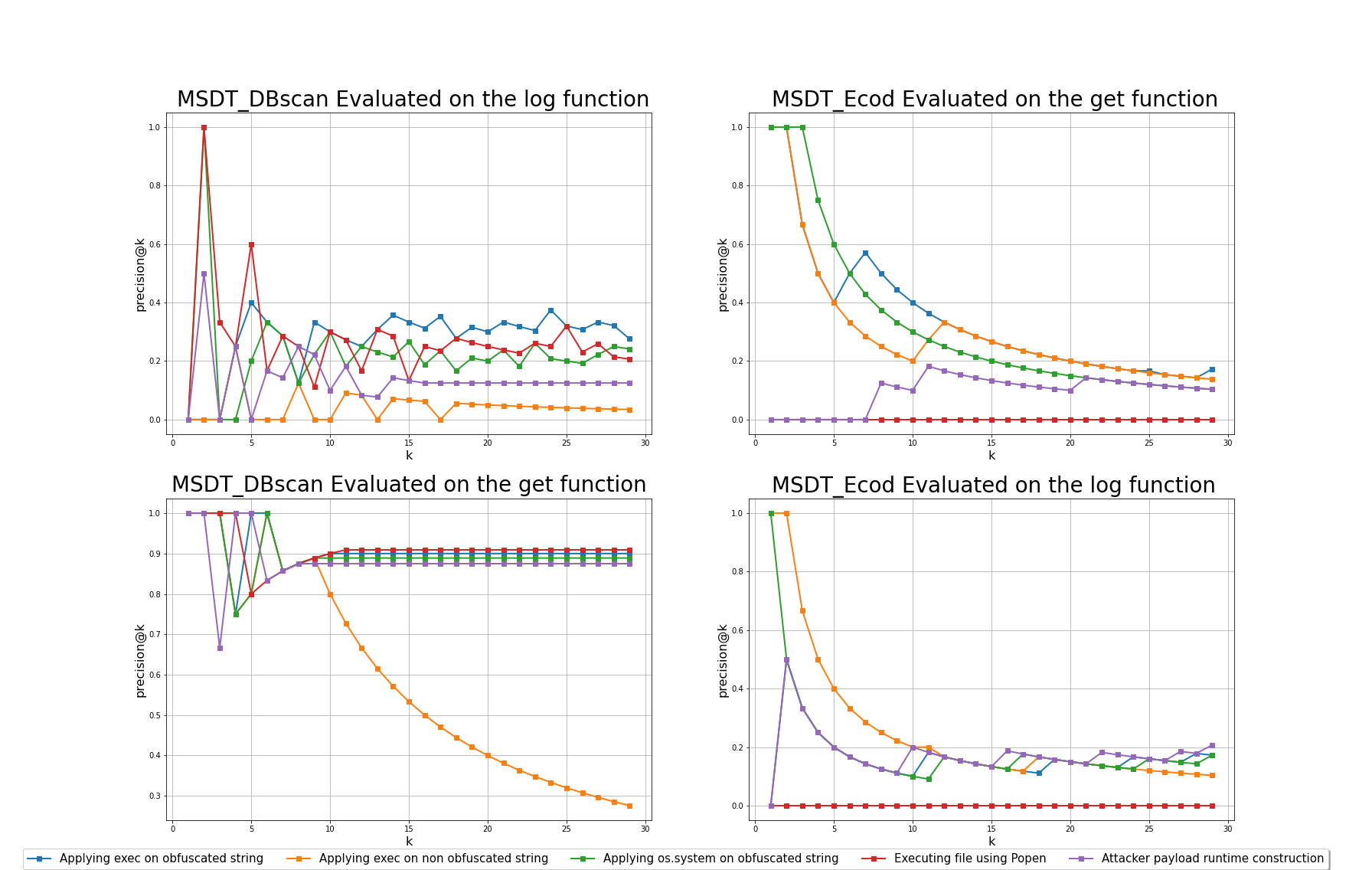}
\caption{\label{fig:get_log_at_k} The measured \textit{precision@k} of $MSDT_{DBSCAN}$ and $MSDT_{Ecod}$ of the \textit{get} and the \textit{log} functions' implementations.}
\end{figure*}

The second experiment included the evaluation of $MSDT_{DBSCAN}$ on every function type against every attack type and every $k$ in the range of 1 to 10 percent of the implementations. For every iteration of $k$, we measured \textit{precision@k}. We found that $MSDT_{DBSCAN}$ manages to detect well when applied to several functions and attacks. Such as the \textit{get} function with three of the mentioned attacks, for $k=10$, MSDT presented the highest value of $precision@10=0.909$ (see Figure~\ref{fig:get_log_at_k}), compared to  $precision@10=0$ which was obtained by the  $Random Classifier$. On the other hand, we found that $MSDT_{DBSCAN}$ achieved less successful results on several functions no matter the type of the applied attack, and the value of the $k$. Such as the \textit{log} function with all the attacks, specifically with the non-obfuscated attack.
Table~\ref{tab:sample_of_tpr_table} and
Appendix~\ref{sec:all_functions_implementations_tpr} present in detail the results of these experiments.

In addition, we discovered that the measured Spearman's rank correlation between the MSDT'S detection rate and the number of implementations is equal to $\rho=0.539$, which indicates a correlation between the detection rate and the number of implementations.

\begin{figure}[!h]
\includegraphics[width=\linewidth]{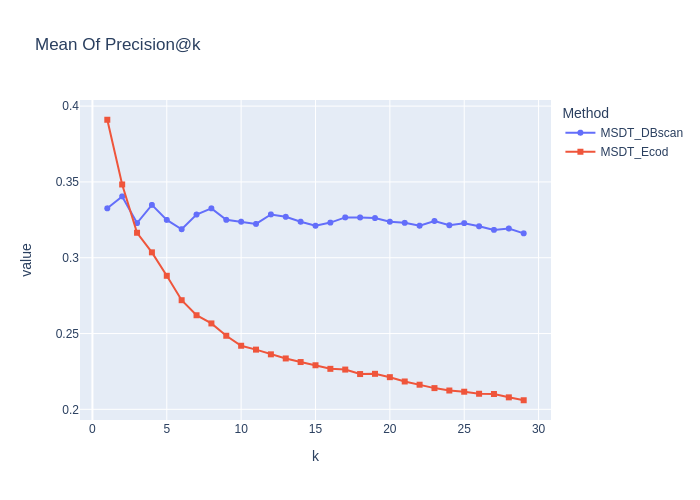}
\caption{\label{fig:mean_precision_k} The measured mean \textit{precision@k} of $MSDT_{DBSCAN}$ and $MSDT_{Ecod}$ of all the 100 function types and the 5 attacks.}
\end{figure}

Additionally, we tested the $MSDT_{Ecod}$ on the same experiment settings described in Section~\ref{sec:code2seq_representation}. Followed by the mentioned evaluation (see Section~\ref{sec:evaluation_metrices}), we measured the \textit{precision@k} for every $k$ in range of 1 to 30. We can observe that generally the $MSDT_{Ecod}$ detects the top 2 rank anomalies, and less successful in the following $k$ values (see Figure~\ref{fig:mean_precision_k}).

\begin{table*}
    \centering
    \caption{\textit{precision@k} for 3 functions with all attacks and $k$ values. The complete \textit{precision@k} results shown in Appendix \ref{chap:appendix}}
    \begin{tabular}{|l|l|p{15mm}p{20mm}p{20mm}p{20mm}p{20mm}p{20mm}|}
    \hline
        \textbf{Model} & \textbf{Function Name} & \textbf{k} & \textbf{Execution
        of an obfuscated string using \textit{exec}} & \textbf{Execution of a non obfuscated script using \textit{exec}} & \textbf{Execution of a obfuscated string using \textit{os.system}} & \textbf{Loading a file from the root directory of the program} & \textbf{Payload construction as an obfuscation use case} \\ \hline
        \multirow{9}{*}{\textbf{$MSDT_{DBSCAN}$}} & \multirow{3}{*}{get} & 10 & 0.9 & 0.8 & 0.889 & 0.9 & 0.7\\ \cline{3-8}
        & & 20 & 0.9 & 0.4 & 0.889 & 0.909 & 0.35 \\ \cline{3-8}
        & & 30 & 0.9 & 0.267 & 0.889 & 0.909 & 0.233 \\ \cline{2-8}
        & \multirow{3}{*}{log} & 10 & 0.4 & 0.1 & 0.4 & 0.3 & 0.3 \\\cline{3-8}
        & & 20 & 0.15 & 0.05 & 0.25 & 0.25 & 0.2 \\\cline{3-8}
        & & 30 & 0.3 & 0.033 & 0.267 & 0.233 & 0.267 \\\cline{2-8}
        & \multirow{3}{*}{update} & 10 & 0.7 & 0.167 & 0.7 & 0.7 & 0.6 \\\cline{3-8}
        & & 20 & 0.733 & 0.167 & 0.722 & 0.75 & 0.706 \\\cline{3-8}
        & & 30 & 0.733 & 0.167 & 0.722 & 0.821 & 0.706 \\ \hline
        \multirow{9}{*}{\textbf{$MSDT_{Ecod}$}} & \multirow{3}{*}{get} & 10 & 0.5 & 0.4 & 0.3 & 0.1 & 0.2 \\ \cline{3-8}
        & & 20 & 0.3 & 0.25 & 0.15 & 0.05 & 0.1 \\ \cline{3-8}
        & & 30 & 0.276 & 0.172 & 0.138 & 0.034 & 0.103 \\ \cline{2-8}
        & \multirow{3}{*}{log} & 10 & 0.3 & 0.1 & 0.1 & 0.2 & 0.2 \\ \cline{3-8}
        & & 20 & 0.15 & 0.15 & 0.1 & 0.1 & 0.2 \\ \cline{3-8}
        & & 30 & 0.172 & 0.103 & 0.103 & 0.069 & 0.172 \\ \cline{2-8}
        & \multirow{3}{*}{update} & 10 & 0.2 & 0.5 & 0.4 & 0.1 & 0.2 \\ \cline{3-8}
        & & 20 & 0.2 & 0.35 & 0.35 & 0.05 & 0.2 \\ \cline{3-8}
        & & 30 & 0.172 & 0.276 & 0.276 & 0.038 & 0.241 \\ \hline
    \end{tabular}
\label{tab:sample_of_tpr_table}
\end{table*}

\begin{figure}[!h]
\includegraphics[width=\linewidth]{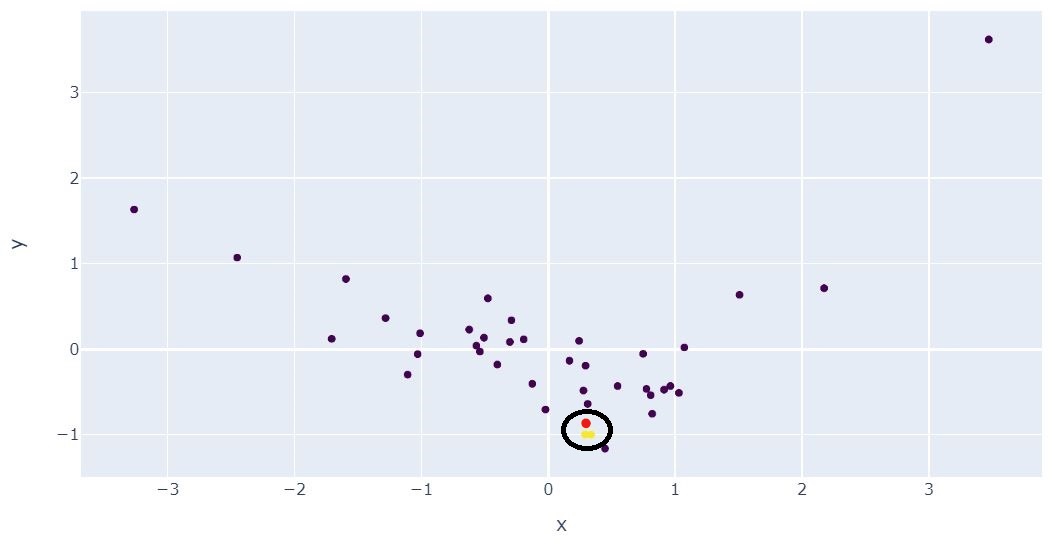}
\caption{\label{fig:multiply}PCA (2 compensates) visualization of real-case detection. The red data point is the attacked function, and the two yellow data points are the unique functions.}
\end{figure}

The third experiment included detecting injected malicious implementations of \textit{multiply} by applying $MSDT_{DBSCAN}$ on it. By visualizing the PCA (2 components) of the collected samples (see Figure~\ref{fig:multiply}), we can see that detecting the attacked functions, for this case, is not a straightforward task. Additionally, we can see (see Figure~\ref{fig:multiply}) that by applying $MSDT_{DBSCAN}$, we managed to detect the malicious implementation, along with two unique and odd implementations\footnote{Those implementations include:(1) Adding in a for loop the first input number by the second input number; and (2) Output the result by comparing the two input number to a results dictionary.} of \textit{multiply}. Then we compared the results of this experiment to \textit{Bandit} and \textit{Snyk} \ref{sec:static_analysis}, yielding that those static analysis tools failed to detect these attacks. Additionally, we compared $MSDT_{DBSCAN}$ to $MSDT_{Ecod}$, which detects only one of the mentioned unique implementation.

\begin{figure}[!h]
\centering
\includegraphics[width=0.5\textwidth]{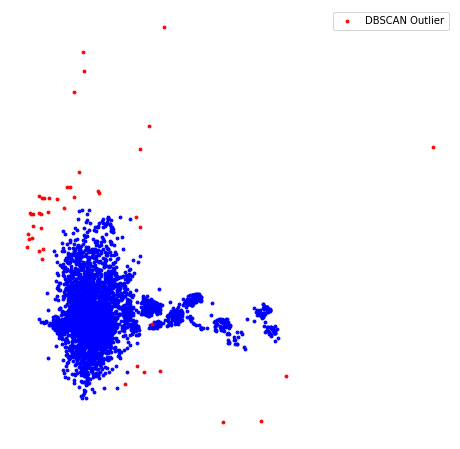}
\caption{\label{fig:dbscan_get} PCA of the \textit{get} function benign (blue) and malicious (red) implementations.}
\end{figure}

\begin{figure}
\centering
\includegraphics[width=0.5\textwidth]{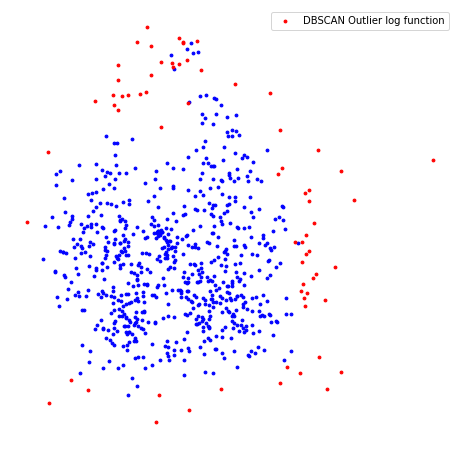}
\caption{\label{fig:dbscan_log} PCA of the \textit{log} function benign (blue) and malicious (red) implementations.}
\end{figure}

The fourth experiment emphasizes the relations between malicious and benign implementations. By the following visualization we received (see Figures~\ref{fig:dbscan_get} and \ref{fig:dbscan_log}) that the \textit{get} functions tend to cluster and on the other hand \textit{log} functions do not cluster well. Therefore, this illustrates the differences in the distribution of the various function types.

%% file: discussions.tex
\label{chap:discussions}
Based on our analysis of the results presented in Section~\ref{chap:results} and Appendix~\ref{sec:all_functions_implementations_tpr}, we can observe the following: 

First, $MSDT_{DBSCAN}$, which detects malicious code injections to functions by anomaly detection on an embedding layer, had promising results when evaluated on different function types with various injected attacks, reaching to \textit{precision@k} up to 0.909 with median=0.889 and mean=0.807 for \textit{get} and \textit{list} function types (see Appendix \ref{sec:all_functions_implementations_tpr} and Figure \ref{fig:get_log_at_k}).

Second, $MSDT_{DBSCAN}$ achieved successful compared to other tools and methods (see Table \ref{tab:sample_of_tpr_table} and Figure \ref{fig:mean_precision_k}). For example, the general \textit{precision@k} of $MSDT_{DBSCAN}$ is higher for $k>2$ compare to the $MSDT_{Ecod}$ based method (as can seen in Section~\ref{fig:mean_precision_k}).
As mentioned in Section~\ref{sec:injection_simulation} the simulated injections are taken from real-world cases and injected into functions. For illustrating a real-world code injection detection we conducted an empirical experiment, which includes detecting real-world attack by $MSDT_{DBSCAN}$ (see Section~\ref{sec:evaluation_metrices}). We got that $MSDT_{DBSCAN}$ results seem promising compared to other widely use static analysis tools and $MSDT_{Ecod}$, in this specific case (see Figure \ref{fig:multiply} and Section \ref{chap:results}). In the future, we would evaluate $MSDT_{DBSCAN}$ on other real-world cases and test on different Program Language functions.
In addition, we can notice that the mentioned static analysis tools are only able to work on files whilst MSDT works on functions. On the one hand, this gives a more precise ability to detect code injections to functions. On the other hand, when applied to rare functions without many implementations, MSDT would not necessarily succeed. In this case, we would like to test whether applying MSDT on similar functions helps to detect code injection in rare functions.

Third, we observed that when $MSDT_{DBSCAN}$ evaluated on similar attacks we get similar results. For example the attacks that use \textit{exec} and \textit{os.system} (as can seen in \textit{get} results in Figure \ref{fig:get_log_at_k}) using the same payload but different execution functions. Additionally, we can see that the \textit{precision@k} values is relatively similar for these two attacks in general (see Appendix \ref{sec:all_functions_implementations_tpr}). This conclusion shows us that if $MSDT_{DBSCAN}$ manages to detect some attack well then it should detect another semantically related attack - we would like to explore this further in future works.

Fourth, we found that $MSDT_{DBSCAN}$ seems to succeed when applied to functions with specific functionality that repeats in the various implementations of the same function type. For example, the \textit{update} implementations tends to be similar - in general this type of function gets an object and calculates or gets as an input a new value to insert in the given object  - as we can see in Appendix \ref{sec:all_functions_implementations_tpr} for functions like \textit{reset}, \textit{list}, and \textit{update} are with a main functionality and a relatively high \textit{precision@k}. In this case, the various implementations of the same function type are similar semantically, yielding that the embedding for each of those is close, hence cluster well (see Figure \ref{fig:dbscan_get} for illustration).

Fifth, we found that $MSDT_{DBSCAN}$' detection rate is positively correlated to the number of implementations in the function type. Hence, $MSDT_{DBSCAN}$ is more likely to achieve a higher detection rate with a more common function type with numerous implementations.

Sixth, when injecting attacks with large line lengths, such as the non-obfuscated script execution, $MSDT_{DBSCAN}$ tends to achieve less successful results (see Figure \ref{fig:get_log_at_k}). For example when evaluating $MSDT_{DBSCAN}$ on the different function types injected with the non obfuscated script, we generally get a low \textit{precision@k} (see Appendix \ref{sec:all_functions_implementations_tpr}). In this case, the injected functionality is a script with numerous lines, which probably affects the Code2Seq robustness and causes it to miss-infer the function's functionality, as researched by Ramakrishnan et al. \cite{ramakrishnan2020semantic}. In future work, we would like to create with Code2Seq and a more robust model for source code (such as Seq2Seq \cite{ramakrishnan2020semantic}), stacking model to overcome Code2Seq vulnerabilities.

Seventh, we can observe that $MSDT_{DBSCAN}$ tended to achieved less successful results when applied on abstract functions with functionality that does not repeat in other implementations - as we can see in the Appendix \ref{sec:all_functions_implementations_tpr} for functions like \textit{run}, \textit{main} etc. For example \textit{install} function, generally, this function is supposed to change the state of the endpoint by activities that belong to the installation process (each application has a different process), such as writing files to disk or establishing a connection with a remote server, etc. Each application has a different process with its unique activities to install the app. In this case, the various implementations of the same function type are inherently different, yielding that the embedding for each of those is not close, hence does not cluster well (see Figure \ref{fig:dbscan_log} for illustration). 
However, we will able to detect anomalies with $MSDT_{DBSCAN}$ with given versions of the abstract function.

Finally, as can observe from the results, statically detecting code injection within functions is a difficult and not homogeneous task for all of the various cases, such as function and attack types. However, MSDT had shown successful results for some cases simulated in the experiments. Therefore MSDT can be used as a detection tool that indicates what function need further investigation, thus reducing the search space and allowing prioritizing anomalies.

%% file: conclusion_and_future_work.tex
\label{chap:conclusion}
This study introduces MSDT, a novel algorithm to statically detect code injection in functions' source code by utilizing a transformer-based model named Code2Seq, and applying anomaly detection techniques on Code2Seq's representation for each function type. We provided a comprehensive description of MSDT’s steps, which start with a collection of a dataset and preprocessing it. After injecting five malicious functionalities into random implementations, we extracted embedding for each one of the implementations in the function type. Based on these embeddings, we managed to apply an anomaly detection technique, resulting in anomalies that we eventually ranked by their distance from the nearest cluster border point.

This evaluation of MSDT on the constructed dataset demonstrates that MSDT succeeded for cases when: (1) The functions have a repetitive functionality; and (2) The injected code has a limited number of lines. However, MSDT was less successful when: (1) The injected code contains a relatively large number of lines; and (2) The functions have a more abstract functionality.

For the MSDT to use the Code2Seq embedding, it is necessary to convert every function to an AST representation. A possible future research direction is using a more comprehensive representation for a code that includes the semantic, syntactic, and execution flow data of the program. For instance, using execution paths in a control flow graph \cite{alomari2019scalable, yu2019empirical} that have been constructed statically from a program. Another possible research direction can be exploring other models than Code2Seq for source code embeddings, like Seq2Seq, CodeBERT, and CodeX.

Those future works are direct conclusions from the MSDT evaluation and results. Therefore, we believe that this future research along with MSDT can create more secure software products and more effective software development procedures.

%% file: data_and_code_availability.tex
\label{chap:data_and_code_availability}
The code that implements our simulations (see Section \ref{sec:injection_simulation}) and the simulated datasets we created (see Section \ref{sec:datasets}) will be available after publication upon request.

%% file: appendix.tex
\label{chap:appendix}

\includepdf[pages=1,scale=0.9,offset=0mm -75,pagecommand={
  \begin{flushleft}
  \begin{appendices}
    \section{All functions implementations TPR}
    \label{sec:all_functions_implementations_tpr}
    The graphs below describe the \textit{precision@k} results of the applied method in $k$ in the range of 1 to 30. The presented results including all function implementations with different attacks (with the random code injection, see Section \ref{sec:injection_simulation}).
    \end{appendices}
  \end{flushleft}}]{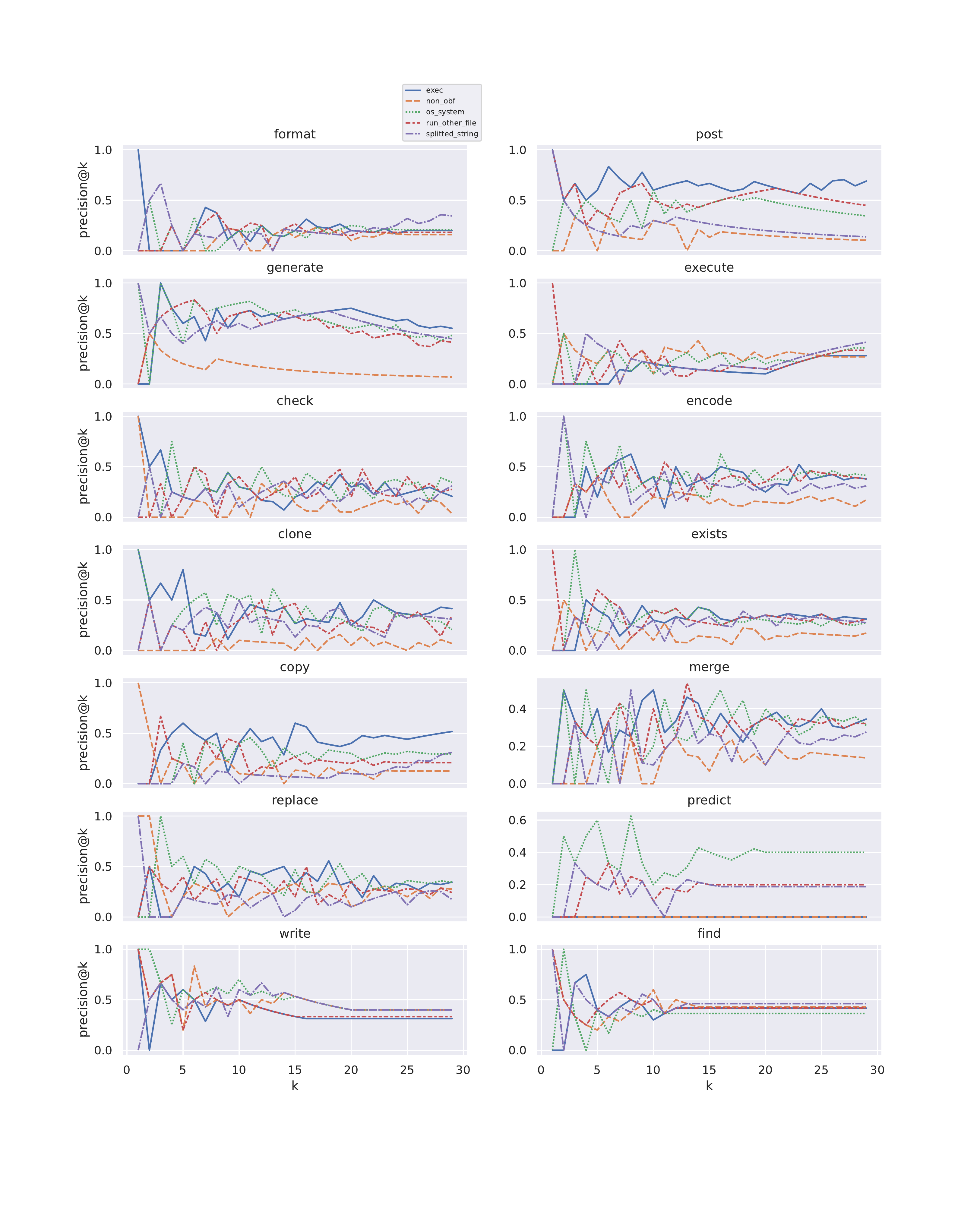}
\includepdf[pages={2-},scale=1]{plots_subplots_final.pdf}